\documentclass[paper, long, noblind]{geophysics}

\newcommand\given[1][]{\,#1\lvert\,}

\usepackage[T1]{fontenc}
\usepackage[utf8]{inputenc}
\usepackage{url}
\usepackage{amsmath, amssymb}
\usepackage[hidelinks]{hyperref}
\usepackage{graphicx}
\usepackage{subfig}
\usepackage{enumitem}
\usepackage[disable]{endfloat}
\usepackage{tabularx}
\usepackage{pdfpages}

\AtBeginDelayedFloats{\linespread{2.0}}

\hypersetup{
    pdftitle={Generative spectral IP modeling},
    pdfauthor={Charles L. Bérubé},
    bookmarksnumbered=true,
    bookmarksopen=true,
    bookmarksopenlevel=1,
    colorlinks=false,
    allcolors=blue,
    pdfstartview=Fit,
    pdfpagemode=UseOutlines,    
    pdfpagelayout=TwoPageRight
}

\begin{document}

\includepdf{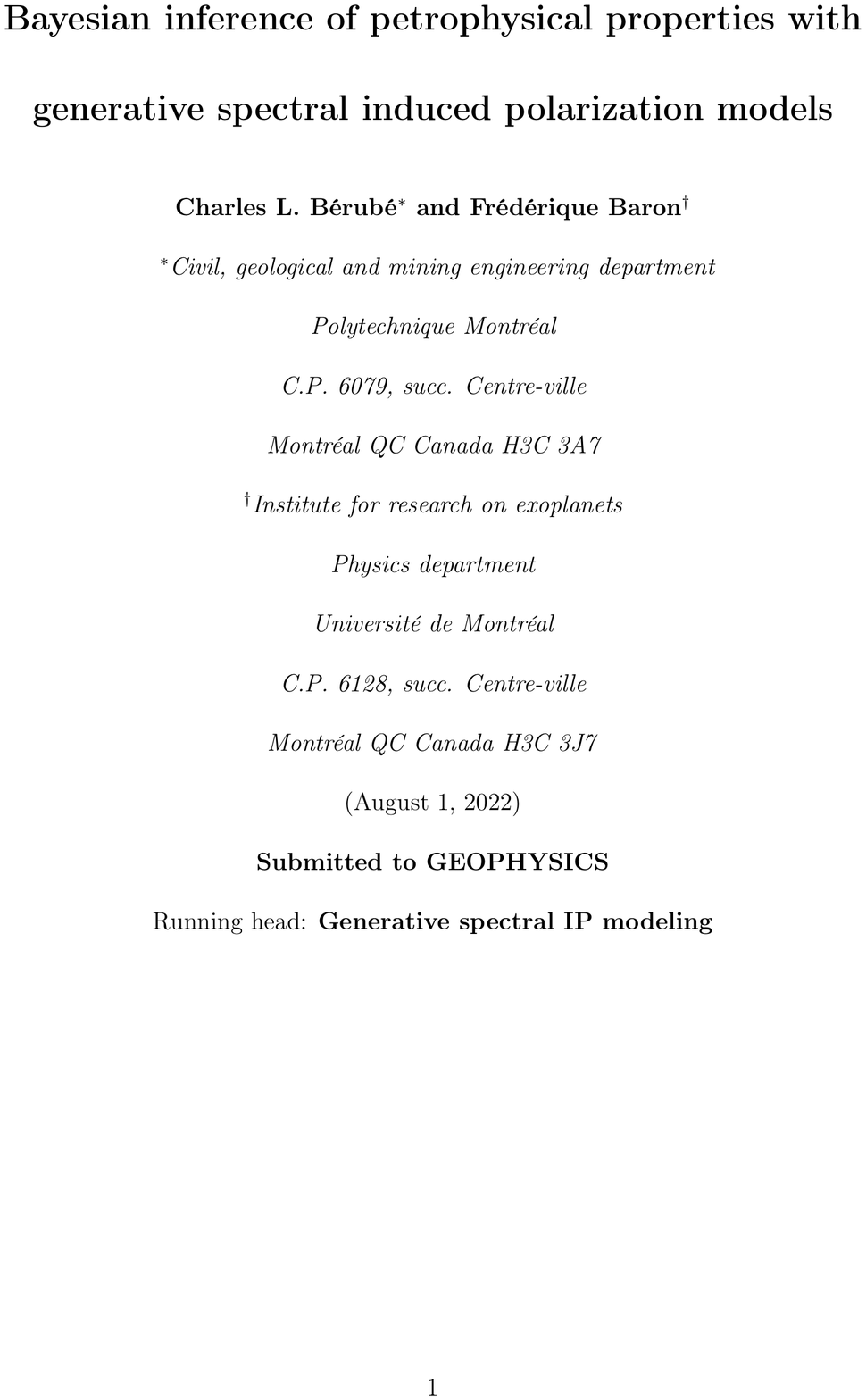}

\title{Bayesian inference of petrophysical properties with generative spectral induced polarization models}

\renewcommand{\thefootnote}{\fnsymbol{footnote}}

\ms{Submitted to GEOPHYSICS} 

\address{
\footnotemark[1]Civil, geological and mining engineering department\\
Polytechnique Montréal\\
C.P. 6079, succ. Centre-ville\\
Montréal QC Canada H3C 3A7 \\
\footnotemark[2]Institute for research on exoplanets\\
Physics department\\
Université de Montréal\\
C.P. 6128, succ. Centre-ville\\
Montréal QC Canada H3C 3J7}
\author{Charles L. Bérubé\footnotemark[1] and Frédérique Baron\footnotemark[2]}

\footer{Submitted to GEOPHYSICS}
\lefthead{Bérubé \& Baron}
\righthead{Generative spectral IP modeling}

\maketitle

\begin{abstract}
Mechanistic induced polarization (IP) models describe the relationships between the intrinsic properties of geomaterials and their frequency-dependent complex conductivity spectra. However, the uncertainties associated with estimating petrophysical properties from IP data are still poorly understood. Therefore, practitioners rarely use mechanistic models to interpret actual IP data. We propose a framework for critically assessing any IP model's sensitivity and parameter estimation limitations. The framework consists of a conditional variational autoencoder (CVAE), an unsupervised Bayesian neural network specializing in data dimension reduction and generative modeling. We train the CVAE on the IP signatures of synthetic mixtures of metallic mineral inclusions in electrolyte-filled host geomaterials and describe the effect of data transformations on the model. First, the CVAE's Jacobian reveals the relative importance of each petrophysical property for generating spectral IP data. The most critical parameters are the conductivity of the host, the volumetric content of the inclusions, the characteristic length of the inclusions, and the permittivity of the host. The inclusions' diffusion coefficient, permittivity, and conductivity, as well as the host's diffusion coefficient, only have marginal importance for generative IP modeling. A parameter estimation experiment yields the standardized accuracy of petrophysical properties using various model constraints scenarios and corroborates the sensitivity analysis results. Finally, we visualize the effects of data transformations and model constraints on the petrophysical parameter space. We conclude that a common logarithm data transformation yields optimal parameter estimation results and that constraining the electrochemical properties of the geomaterial improves estimates of the characteristic length of its metallic inclusions and vice versa.
\end{abstract}

\section{Introduction}
A wide range of geoscience and engineering applications rely on the induced polarization (IP) method to characterize the electrical properties of the subsurface. The IP effect refers to the subsurface's temporary and reversible energy storage, which happens when porous geomaterials become polarized under the influence of a transient electrical field. A variety of mechanisms govern the electrical polarization of geomaterials.
However, the polarization of the electrical double layer at the interfaces between metallic particles and interstitial electrolytes is known to cause some of the most important measurable IP effects, as demonstrated by numerous laboratory experiments on synthetic geomaterials \citep[e.g.,][]{scott_induced_1969, mahan_complex_1986, gurin_induced_2015, mao_induced_2016, gurin_spectral_2021}. Therefore, characterizing buried mineral deposits has historically been one of the main driving factors for developing the IP method. Nevertheless, metallic particles are not the only source of polarization in geomaterials. Soil clay content, cation exchange capacity, and membrane polarization cause measurable IP effects. Researchers are thus expanding the applications of the IP method to hydrogeology \citep{gazoty_application_2012}, shallow landslide risk assessment \citep{revil_induced_2020}, groundwater exploration \citep{azffri_electrical_2022}, and evaluation of organic matter content in agricultural practices \citep{schwartz_spectral_2015}.

It is considerably more challenging to interpret the IP signatures of actual rocks than those of synthetic geomaterials because the former may exhibit complex folding, fracturing, alteration, weathering, and anisotropy. Studies that combine scanning electron microscopy and the IP method show that the intensity of the IP effect may vary with the volumetric content of disseminated conductive minerals in the rocks \citep{johansson_investigations_2017, berube_mineralogical_2018, gurin_induced_2018}, although direct relationships are not evident. Interpreting IP data is even more challenging in the vicinity of ore deposits because alteration products such as microcline, calcite, albite, and carbonate minerals may encapsulate sulfide minerals, effectively preventing the interfacial polarization effect \citep{berube_mineralogical_2018, gurin_induced_2019}.
Nevertheless, the IP method has largely been applied to explore buried graphite, magnetite and pyrrhotite deposits \citep{pelton_mineral_1978}, gold-rich porphyry copper systems \citep{close_electrical_2001}, volcanic massive sulphide deposits \citep{tavakoli_deep_2016}, iron oxide-copper-gold deposits \citep{aguilef_relationship_2017}, and high-volume, low-grade disseminated gold deposits \citep{berube_mineralogical_2018}.
Other applications of the IP method in the mining and energy sectors include evaluating potential resources for mineral reuse in mining slag heaps \citep{gunther_spectral_2016}, monitoring the development of acid mine drainage in mine tailings \citep{placencia-gomez_spectral_2015}, and characterizing pyrite-altered sedimentary rocks above hydrocarbon accumulations \citep{veeken_benefits_2009, veeken_modelling_2012}.

Empirical, mechanistic, or data-driven models are the three main methods for interpreting the polarization of disseminated metallic particles in geomaterials. Empirical models primarily consist of simple equivalent circuits with resistors and capacitors that mimic the polarization response of metallic particles \citep[e.g.,][]{pelton_mineral_1978, dias_developments_2000}.
Mechanistic models, on the other hand, consider the intrinsic properties of the geomaterials and are derived from the underlying physics of the polarization phenomena \citep[e.g.,][]{wong_electrochemical_1979, revil_induced_2015, misra_interfacial_2016, bucker_electrochemical_2018, jin_mechanistic_2019}.
Alternatively, data-driven models aim to capture the discriminant features in IP data without assuming the underlying polarization mechanism. Data-driven IP models include, for example, deep generative neural networks trained on extensive compilations of field data \citep{berube_data-driven_2022}.

Researchers have mostly focused on developing tools to interpret IP data with empirical models. Among these models, the Cole-Cole, Warburg, and Debye decomposition approaches are widely used with either deterministic \citep{nordsiek_new_2008, ustra_relaxation_2016, weigand_debye_2016} or stochastic \citep{ghorbani_bayesian_2007, chen_comparison_2008, keery_markov-chain_2012, berube_bayesian_2017} curve fitting methods to estimate the IP parameters of geomaterials. In general, the empirical parameters are (1) the direct current apparent conductivity, which depends on the porosity of the geomaterial and the conductivity of its saturating fluid, (2) the chargeability, which generally increases with the volumetric content of polarizable minerals \citep{revil_induced_2015, abdulsamad_spectral_2017}, (3) the characteristic relaxation time, which is highly dependent on the grain size of polarizable particles and the saturating electrolyte salinity \citep{gurin_time_2013, abdulsamad_spectral_2017}, and (4) the frequency dependence exponent. The interpretation of the frequency dependence exponent is ambiguous. It can be estimated as a model parameter for Cole-Cole decomposition \citep{chen_comparison_2008}, set to $1$ for Debye decomposition \citep{morgan_inversion_1994}, or set to $1/2$ for Warburg decomposition \citep{revil_spectral_2014}.

Mechanistic models establish direct relationships between IP data and the petrophysical properties of geomaterials. These properties include the intrinsic conductivity, permittivity, and diffusion coefficients of the host material and those of the metallic mineral inclusions. Mechanistic models can also consider the volumetric content and grain size of the inclusions through effective medium theory \citep[e.g.,][]{misra_interfacial_2016, jin_mechanistic_2019}. Interpreting data through mechanistic models is desirable because it would theoretically allow users of the IP method to directly estimate the petrophysical properties using non-invasive observations. However, engineers and practitioners seldom use mechanistic models in real-world applications of the IP method because the mathematical equations describing mechanistic models are much more complex than their empirical counterparts. In addition, model parameter definitions vary depending on the initial assumptions made about the interfacial polarization phenomenon, which hinders the comparison of data sets interpreted with different mechanistic models \citep[e.g.,][]{revil_induced_2015, misra_interfacial_2016, bucker_electrochemical_2018}. Moreover, global sensitivity and uncertainty analyses of mechanistic models’ petrophysical properties are mostly absent from the geophysical literature. The accuracy with which we can estimate these properties from IP data is thus poorly understood. A notable exception is \cite{placencia-gomez_electrochemical_2014}, who specifically analyze the sensitivity of the \cite{wong_electrochemical_1979} model to the oxidation of sulfide minerals. Finally, it is challenging to constrain the intrinsic electrochemical properties of metallic minerals because values reported in the literature may vary across orders of magnitude due to impurities \citep[e.g.,][]{pridmore_electrical_1976, emerson_pyrite_2019}.

In addition to the aforementioned conceptual challenges, there are implementation challenges for fitting mechanistic models to IP data. Spectral IP data are complex-valued, and their in-phase and quadrature components can vary across multiple orders of magnitude. Previous studies on fitting spectral IP data with empirical models consider the in-phase and quadrature components as a concatenation of two real-valued features \citep[e.g., ][]{ghorbani_bayesian_2007, chen_comparison_2008, nordsiek_new_2008, ghorbani_cr1dinv:_2009, weigand_debye_2016, berube_bayesian_2017}. However, with this method, the objective function may favor fitting only one of the components if it is much larger in value than the other. To ensure they have equal importance during optimization, we can individually normalize the in-phase and quadrature components, although this strategy violates the Kramers--Kronig relations. An other approach is to use complex-valued neural networks for curve fitting \citep{virtue_complex-valued_2019}, but most open source deep learning frameworks currently have limited complex numbers support.

The goal of this study is to critically assess the applicability of mechanistic models for IP data interpretation. To do so, we develop a data-driven autoencoder framework to analyze mechanistic IP models' sensitivity and parameter estimation limitations. The framework is applied to the so-called "perfectly polarized interfacial polarization" (PPIP) model, initially proposed in \cite{misra_interfacial_2016} and extended in \cite{jin_mechanistic_2019}. The theoretical contributions of this study are three-fold : (1) we analyze the generative modeling sensitivity of the PPIP model to its petrophysical parameters, (2) we quantify the parameter estimation limitations of this model under various conditioning scenarios, and (3) we compare the effects of diverse data transformation strategies on the model's sensitivity and effective parameter space. Finally, as a practical contribution, we release all the codes required to reproduce our experiments as an open-source Python package to facilitate the future critical assessment of mechanistic IP models. The following two sections define the PPIP model and establish the data-driven framework. The results section describes over-parameterization in the PPIP model and highlights the effects of data transformations on the model's sensitivity, parameter estimation limitations, and effective parameter space.

\section{Polarization of conductive minerals}

\subsection{Qualitative description}
A medium comprising a host phase and an inclusion phase can approximate a geomaterial with disseminated metallic mineral inclusions. The host phase represents the electrolyte solution, which fills the porosity network of the geomaterial and whose charge carriers are cations and anions. The inclusion phase represents metallic (e.g., pyrite or magnetite) or semi-metallic minerals (e.g., graphite). Metallic minerals are semi-conductors whose charge carriers are electrons and holes \citep{revil_induced_2015}.

\begin{figure}
\centering
\subfloat[\label{fig:polarization1}]{%
  \includegraphics[clip, width=0.45\textwidth, page=1]{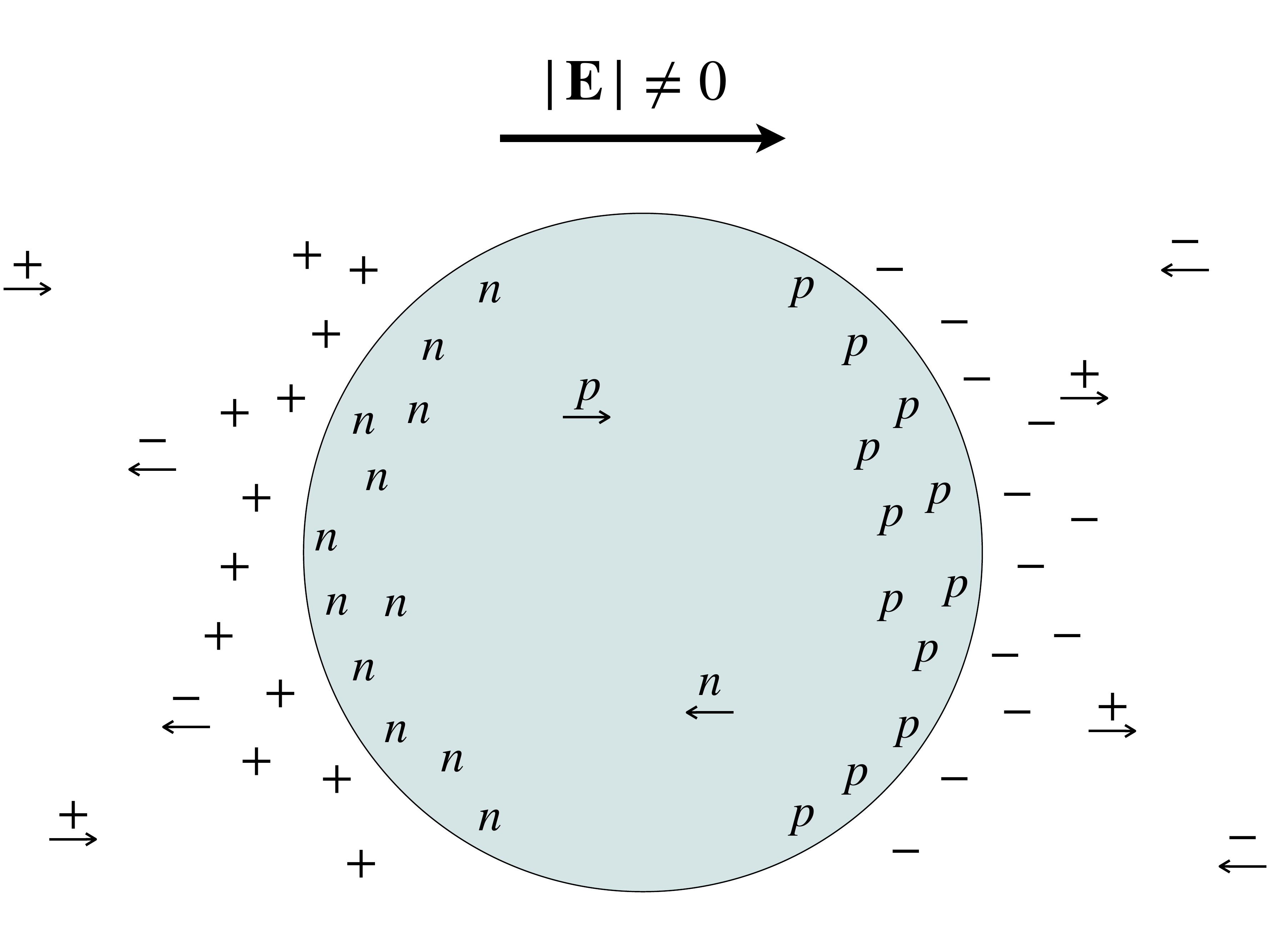}
}
\hfill
\subfloat[\label{fig:polarization2}]{%
  \includegraphics[clip, width=0.45\textwidth, page=2]{figures/polarization}
}
\caption{(a) Interfacial polarization of a spherical metallic mineral inclusion in a host electrolyte solution caused by electromigration of charge carriers in the presence of an applied electric field $\mathbf{E}$. (b) Charge carrier diffusion towards electric charge neutrality in the host and inclusion phases during the relaxation time that follows the removal of the applied electric field. $n$ and $p$ are electrons and holes in the metallic mineral, respectively. In the electrolyte host, $+$ and $-$ represent cations and anions, respectively. Modified from \cite{misra_interfacial_2016}.}
\label{fig:polarization}
\end{figure}

In the presence of an applied electric field, charge carriers within metallic particles migrate toward the interfaces between the host and inclusion phases (Figure~\ref{fig:polarization}). Consequently, holes and electrons accumulate inside the metallic inclusions at the host-inclusion interfaces, whereas cations and anions form a counterion cloud and diffuse layer in the host medium (Figure~\ref{fig:polarization1}). In the absence of redox-active conditions, the transfer of charge carriers from the electrolyte solution to the metallic particles, and vice versa, is not possible \citep{placencia-gomez_electrochemical_2014}. Metallic inclusions are thus considered "perfectly polarized" in the presence of an applied electric field and in the absence of redox-active species \citep{revil_induced_2015, misra_interfacial_2016}.

After removal of the applied electric field, charge carriers diffuse away from the host-inclusion interfaces, and both phases return to electric charge neutrality at a macroscopic scale (Figure~\ref{fig:polarization2}). For metallic inclusions approximated by spherical particles, the characteristic relaxation time with which charge carriers return to equilibrium conditions is directly proportional to the radius of the inclusions at high frequencies and their radius squared at low frequencies \citep{shilov_theory_2010}. Moreover, the characteristic relaxation time is inversely proportional to the logarithm of the diffusion coefficient of charge carriers in the host and inclusion phases \citep{misra_interfacial_2016}. Under a transient applied electric field, geomaterials with metallic mineral inclusions exhibit a frequency-dispersive complex conductivity that depends on electromigration, accumulation, and diffusion processes.

\subsection{Polarizability of spherical inclusions}
\cite{misra_interfacial_2016} provide the mathematical derivation for the polarizability of metallic spherical inclusions in an electrolyte host. The frequency-dependent polarization $f(\omega)$ in the direction normal to the applied electric field for spherical inclusions of radius $a_\mathrm{i}$, charge carrier diffusion coefficient $D_\mathrm{i}$, permittivity $\epsilon_\mathrm{i}$ and intrinsic conductivity $\sigma_\mathrm{i}$ uniformly distributed in a host medium with charge carrier diffusion coefficient $D_\mathrm{h}$, permittivity $\epsilon_\mathrm{h}$ and intrinsic conductivity $\sigma_\mathrm{h}$ is
\begin{equation}
\label{eqn:polarization1}
    f(\omega) = -\frac{1}{2} + \frac{3}{2} \frac{i\omega}{\left[ \frac{2\sigma_\mathrm{h} }{a_\mathrm{i}\epsilon_\mathrm{h}}\frac{E_\mathrm{h}}{G_\mathrm{h}} - \frac{2 K_\mathrm{h}\sigma_\mathrm{i}}{a_\mathrm{i} K_\mathrm{i}\epsilon_\mathrm{i}}\frac{F_\mathrm{i}}{H_\mathrm{i}} + i\omega \left(\frac{2 K_\mathrm{h}}{K_\mathrm{i}} + 1 \right)\right]},
\end{equation}
where $i$ is the imaginary unit and $\omega$ is the angular frequency. $K_\mathrm{i}=\sigma_\mathrm{i}+i\omega\epsilon_\mathrm{i}$ and $K_\mathrm{h}=\sigma_\mathrm{h}+i\omega\epsilon_\mathrm{h}$ are the complex-valued conductivity of the inclusion and host phases, respectively. In Equation~\ref{eqn:polarization1},
\begin{equation}
\label{eqn:polarization2}
    \frac{F_\mathrm{i}}{H_\mathrm{i}} = \frac{a_\mathrm{i}\left[ a_\mathrm{i}\gamma_\mathrm{i} - \tanh{(a_\mathrm{i}\gamma_\mathrm{i})} \right]}{2 a_\mathrm{i}\gamma_\mathrm{i} - (a_\mathrm{i}\gamma_\mathrm{i})^2 \tanh{(a_\mathrm{i}\gamma_\mathrm{i})} - 2\tanh{(a_\mathrm{i}\gamma_\mathrm{i})}}
\end{equation}
and
\begin{equation}
\label{eqn:polarization3}
    \frac{E_\mathrm{h}}{G_\mathrm{h}} = \frac{a_\mathrm{i}(a_\mathrm{i}\gamma_\mathrm{h} + 1)}{(a_\mathrm{i}\gamma_\mathrm{h})^2 + 2 a_\mathrm{i}\gamma_\mathrm{h} + 2},
\end{equation}
where
\begin{equation}
\label{eqn:gamma-i}
    \gamma_\mathrm{i}^2 = \left( \frac{i\omega}{D_\mathrm{i}} + \frac{\sigma_\mathrm{i}}{\epsilon_\mathrm{i}D_\mathrm{i}} \right)
\end{equation}
and
\begin{equation}
\label{eqn:gamma-h}
    \gamma_\mathrm{h}^2 = \left( \frac{i\omega}{D_\mathrm{h}} + \frac{\sigma_\mathrm{h}}{\epsilon_\mathrm{h}D_\mathrm{h}} \right).
\end{equation}
Equations~\ref{eqn:polarization2}~and~\ref{eqn:polarization3} have been reformulated from the equations of $F_\mathrm{i}$, $H_\mathrm{i}$, $E_\mathrm{h}$, and $G_\mathrm{h}$ given in \cite{misra_interfacial_2016} to improve the numerical stability of the PPIP model.

\subsection{Effective medium approximation}
Equation~\ref{eqn:polarization1} describes the microscopic electrical properties of the inclusions. When the observation scale is sufficiently large, the macroscopic electrical properties of a mixture of homogeneous media are approximable through effective medium theory. Under the Maxwell-Garnet approximation, the effective frequency-dependent, complex-valued conductivity ($\sigma_\mathrm{eff}$) of a mixture of spherical metallic inclusions in a host medium filled with an electrolyte solution is
\begin{equation}
    \sigma_\mathrm{eff} = \sigma_\mathrm{h}  \frac{n\phi_\mathrm{i}f(\omega) + 1}{\left[1 - \phi_\mathrm{i} f(\omega)\right]},
\label{eqn:effective-conductivity}
\end{equation}
where $\phi_\mathrm{i}$ is the volumetric content of metallic particles in the mixture, and $n$ is a geometrical factor ($n=2$ for spherical inclusions). $\sigma_\mathrm{eff}$ can be expressed in rectangular form notation by $\sigma_\mathrm{eff}=\sigma'+i\sigma''$, where $\sigma'$ and $\sigma''$ are respectively its in-phase and quadrature components, or in polar form notation by $\sigma_\mathrm{eff}=\vert\sigma_\mathrm{eff}\vert\exp(i\varphi)$, where $\vert\sigma_\mathrm{eff}\vert$ and $\varphi$ are respectively its amplitude and phase angle. The Maxwell-Garnet approximation is valid under the following conditions \citep{jin_mechanistic_2019}: (1) the volume fraction of conductive minerals should be less than 20~\%, (2) the electromagnetic interaction between conductive minerals and other geomaterial components is negligible, (3) the conductive minerals are not in contact with each other, (4) the size of geomaterial heterogeneities is much smaller than the wavelength of the applied electric field, and (5) the skin depth of the electric field in the conductive minerals should not be much smaller than the conductive minerals characteristic length.

\section{Methods}

\subsection{Forward modeling}
This study considers simple mixtures of spherical metallic inclusions disseminated in host electrolyte solutions. For these mixtures, $\sigma_\mathrm{eff}$ depends on eight parameters denoted by $\boldsymbol{\eta} = (a_\mathrm{i}, \phi_\mathrm{i}, D_\mathrm{i}, \sigma_\mathrm{i}, \epsilon_\mathrm{i}, D_\mathrm{h}, \sigma_\mathrm{h}, \epsilon_\mathrm{h})$.
The geometrical properties of the inclusions are described by $a_\mathrm{i}$ and $\phi_\mathrm{i}$, whereas $D_\mathrm{i}$, $\sigma_\mathrm{i}$, and $\epsilon_\mathrm{i}$ are the electrochemical properties of the metallic inclusions, and $D_\mathrm{h}$, $\sigma_\mathrm{h}$, and $\epsilon_\mathrm{r}$ are the electrochemical properties of the host. We generate 100~000 synthetic mixtures by sampling $\boldsymbol{\eta}$ using the Latin hypercube sampling method implemented in SALib \citep{herman_salib_2017}. Table~\ref{tab:parameter-bounds} provides the sampling bounds for each parameter in $\boldsymbol{\eta}$. The parameter bounds are within the limitations of the PPIP model \citep{misra_interfacial_2016, jin_mechanistic_2019}.

\begin{table}
\centering
\caption{Parameter bounds used with the Latin hypercube sampling scheme to generate complex conductivity spectra using the PPIP model and parameters of the test mixture. $\epsilon_0=8.854\times 10^{-12}$~F/m is the permittivity of free space.}
\label{tab:parameter-bounds}
\begin{tabular}{ccrrr}
\hline
Parameter & Units & Minimum & Maximum & Test mixture \\
\hline
$a_\mathrm{i}$ & m & $10^{-5}$ & $10^{-3}$ & $2\times 10^{-4}$ \\
$\phi_\mathrm{i}$ &  & $10^{-3}$ & $0.2$ & $0.1$ \\
$D_\mathrm{i}$ & m\textsuperscript{2}/s & $10^{-7}$ & $10^{-5}$ & $10^{-6}$ \\
$\sigma_\mathrm{i}$ & S/m & $1$ & $10^{5}$ & $10^{2}$ \\
$\epsilon_\mathrm{i}$ & F/m & $10^{-11}$ & $10^{-9}$ & $10\times\epsilon_0$ \\
$D_\mathrm{h}$ & m\textsuperscript{2}/s & $10^{-10}$ & $10^{-8}$ & $10^{-9}$ \\
$\sigma_\mathrm{h}$ & S/m & $10^{-3}$ & $1$ & $10^{-1}$ \\
$\epsilon_\mathrm{h}$ & F/m & $10^{-11}$ & $10^{-9}$ & $80\times\epsilon_0$ \\
\hline
\end{tabular}
\end{table}

Parameter sampling is done in the log space because electrochemical and geometrical properties vary across several orders of magnitude. Next, we randomly split the mixtures into a training data set ($\mathcal{D}_\mathrm{t}$) comprising 80~000 mixtures and a validation data set ($\mathcal{D}_\mathrm{v}$) containing 20~000 mixtures. Finally, the $\sigma_\mathrm{eff}$ of each mixture is computed at 32 logarithmically spaced frequencies in the 100 Hz to 1 MHz range by combining Equations~\ref{eqn:polarization1}~and~\ref{eqn:effective-conductivity}.

\subsection{Data transformations}
We compare the effects of four data transformations on the sensitivity and parameter estimation properties of the PPIP model. Each transformation $T$ maps $\sigma_\mathrm{eff}$ to a concatenation of two real-valued quantities $x'$ and $x''$, as in
\begin{equation}
    T\colon \sigma_\mathrm{eff} \mapsto \left[x', x''\right].
\label{eqn:transforms}
\end{equation}

The first transformation consists of concatenating the raw in-phase and quadrature components of $\sigma_\mathrm{eff}$. This transformation thus reads
\begin{equation}
    T_\mathrm{raw}\colon \sigma_\mathrm{eff} \mapsto \left[\sigma', \sigma''\right],
\label{eqn:raw}
\end{equation}
which preserves the phase angle of complex conductivity.

The second transformation is a common logarithm operation applied individually to the in-phase and quadrature components. The common logarithm transformation is
\begin{equation}
    T_\mathrm{log}: \sigma_\mathrm{eff} \mapsto \left[\log_{10} \sigma', \log_{10} \sigma''\right],
\label{eqn:log}
\end{equation}
which does not preserve the phase angle between both components but prevents them from varying across several orders of magnitude.

The third transformation normalizes both the in-phase and quadrature components in the closed unit interval $[0, 1]$. The normalization transformation reads
\begin{equation}
    T_\mathrm{norm}: \sigma_\mathrm{eff} \mapsto \left[\frac{\sigma' - \min{\sigma'}}{\max{\sigma'} - \min{\sigma'}}, \frac{\sigma'' - \min{\sigma''}}{\max{\sigma''} - \min{\sigma''}}\right].
\label{eqn:normalization}
\end{equation}
Normalization is a widely used technique to ensure equal importance of the input features and to allow faster optimization of a neural network's weights \citep{goodfellow_deep_2016}. However, normalizing the $\sigma_\mathrm{eff}$ components does not preserve the phase angle of the complex conductivity, and information about the relative amplitude of the in-phase and quadrature components is lost.

The fourth transformation consists of computing the principal value of the complex logarithm of $\sigma_\mathrm{eff}$. The principal value of the logarithm of a complex number $c = a + ib$ is defined as
\begin{equation}
    \mathrm{pv}\log \left(c\right) = \ln \vert c\vert + i\arctan{\left(\frac{b}{a}\right)},
\label{eqn:pv-log-def}
\end{equation}
where $\vert c\vert$ is the amplitude of $c$. The components of $\mathrm{pv}\log \left(c\right)$ are then concatenated as real-valued vectors. The principal value transformation for complex conductivity data reads
\begin{equation}
    T_\mathrm{pv}\colon \sigma_\mathrm{eff} \mapsto \left[\ln \vert\sigma_\mathrm{eff}\vert, \varphi\right],
\label{eqn:pv-log}
\end{equation}
which prevents the amplitude from varying across several orders of magnitude and explicitly preserves the phase angle of complex conductivity.

\begin{figure}
\centering
\includegraphics[]{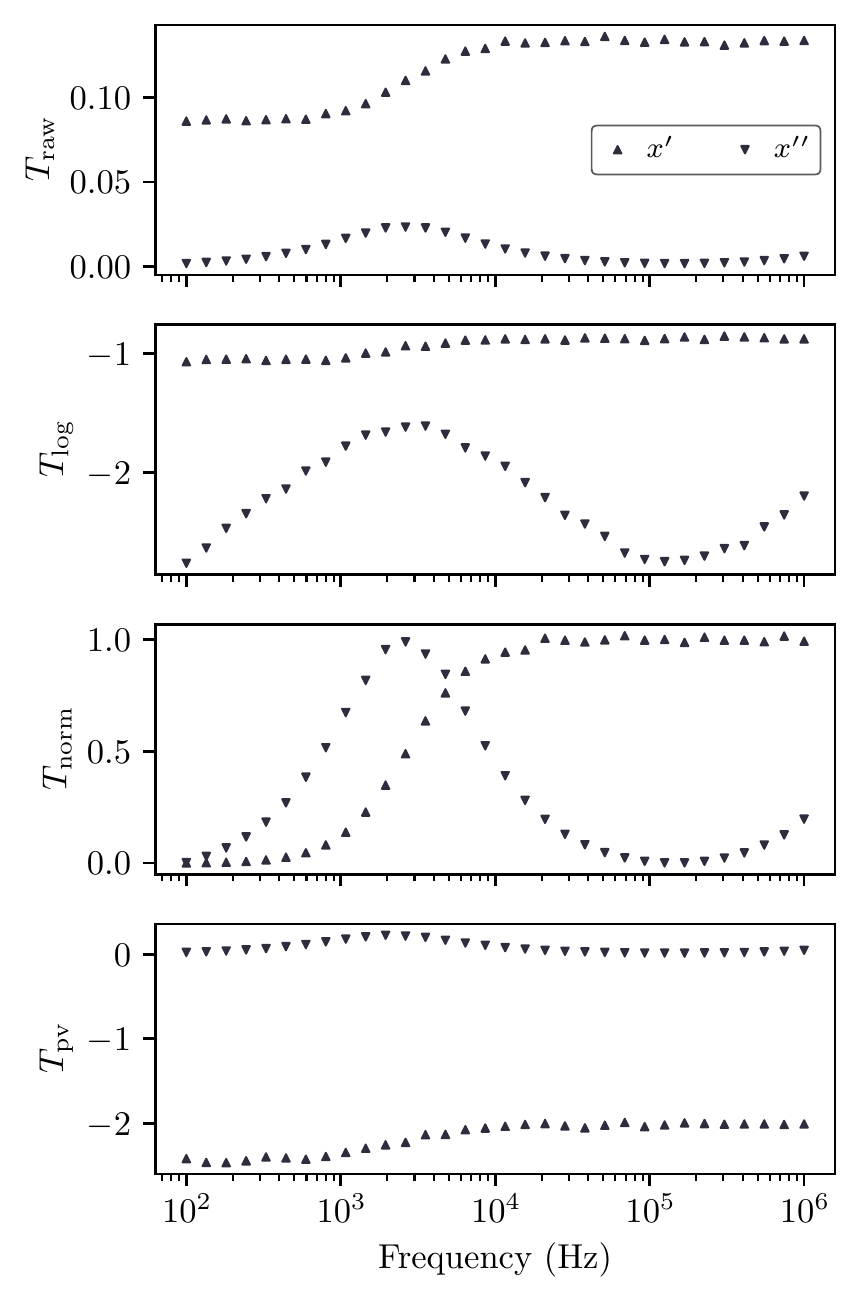}
\caption{Transformed $\sigma_\mathrm{eff}$ spectra ($x'$ and $x''$) of a test mixture of metallic inclusions in a host electrolyte solution. Table~\ref{tab:parameter-bounds} summarizes the test mixture's parameters. One percent additive Gaussian white noise contaminates the data.}
\label{fig:default-mixture-data}
\end{figure}

Figure~\ref{fig:default-mixture-data} shows the result of each data transformation on the complex conductivity spectra of a test mixture whose parameters are given in Table~\ref{tab:parameter-bounds}.

\subsection{Conditional variational autoencoder}
Autoencoders are unsupervised neural networks designed for data dimensionality reduction \citep{hinton_reducing_2006}. Autoencoders comprise an encoder module, which compresses input data ($\mathbf{x}$) to latent representations ($\mathbf{z}$) in a lower-dimensional space, and a decoder module, which expands the latent representations and outputs reconstructions of the data ($\mathbf{\hat{x}}$). Optimization of autoencoders relies on minimizing the reconstruction error between input and output data and is typically achieved with a variant of the stochastic gradient descent algorithm.

Variational autoencoders are generative latent variable models that aim to learn the joint probability $p_\theta(\mathbf{x}, \mathbf{z})$ of observations and latent variables. The posterior distribution $p_{\theta}({\mathbf{z}\given \mathbf{x}})$ is given by Bayes' theorem,
\begin{equation}
    p_{\theta}({\mathbf{z}\given \mathbf{x}}) = \frac{{p_{\theta}( {\mathbf{x}\given \mathbf{z}})\, p_{\theta}(\mathbf{z})}}{{p_{\theta}(\mathbf{x})}},
\label{eq:bayes}
\end{equation}
where $p_{\theta}(\mathbf{z})$ is the prior distribution of the latent vectors, $p_{\theta}( {\mathbf{x}\given \mathbf{z}})$ is the data likelihood, and $p_\theta(\mathbf{x})$ is the marginal likelihood. In practice, the marginal likelihood
\begin{equation}
    p_\theta(\mathbf{x}) = \int p_\theta(\mathbf{x}\given\mathbf{z})\,p_\theta(\mathbf{z})\,\mathrm{d}\mathbf{z},
\end{equation}
is intractable. The posterior distribution can instead be approximated by an encoder neural network $q_{\phi}({\mathbf{z}\given \mathbf{x}})$, parameterized by a set of internal weights $\phi$, whereas $p_{\theta}( {\mathbf{x}\given \mathbf{z}})$ is modeled by a decoder neural network, parameterized by $\theta$ \citep{kingma_auto-encoding_2014}. The optimization objective for variational encoders is the sum of the data reconstruction error and the Kullback-Leibler divergence ($D_\textrm{KL}$) between the latent distribution and a prior on the latent variables. The prior is typically assumed to be the standard normal distribution. Conditional variational autoencoders (CVAE) further expand the concept by integrating prior information about the training data to condition the model. The training objective for a CVAE \citep{sohn_learning_2015} is to maximize the conditional evidence lower bound (ELBO):
\begin{equation}
\label{eqn:elbo}
    \mathrm{ELBO}(\mathbf{x}, \mathbf{c}) = \mathbb{E}\log p_{\theta}\left( \mathbf{x}\given\mathbf{z}, \mathbf{c} \right) - D_\mathrm{KL}\left( {q_{\phi}\left( \mathbf{z}\given\mathbf{x}, \mathbf{c} \right)\parallel p_{\theta}\left( \mathbf{z}, \mathbf{c} \right)} \right),
\end{equation}
where $\mathbf{c}$ are the conditions. In Equation~\ref{eqn:elbo}, the $D_\textrm{KL}$ term acts as a regularization term, whereas the expected log-likelihood term aims to minimize the observed and predicted data misfit.

\subsection{Implementation details}
\begin{figure}
\centering
\includegraphics[width=0.9\textwidth]{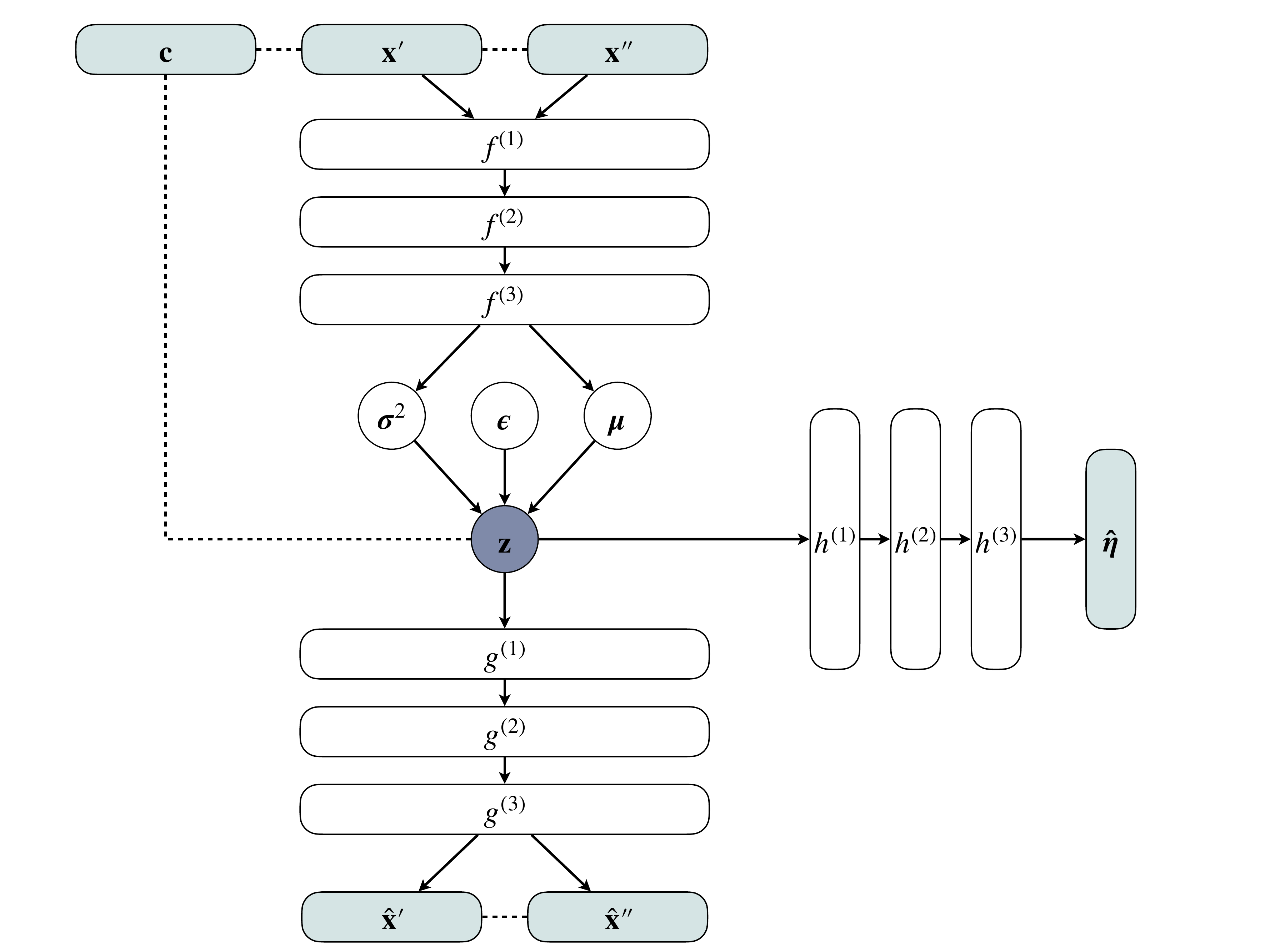}
\caption{Architecture of the proposed CVAE framework. $\mathbf{x'}$ and $\mathbf{x''}$ are the results of the data transformations applied to the effective complex conductivity spectra. $\mathbf{\hat{x}'}$ and $\mathbf{\hat{x}''}$ are reconstructions of the transformed complex conductivity spectra. $\mathbf{c}$ is the conditioning vector (known geomaterial parameters). $\boldsymbol{\hat{\eta}}$ are the predicted geomaterial parameters. $\mathbf{z}$ is a latent distribution sample drawn through the reparameterization trick ($\mathbf{z}=\boldsymbol{\mu} + \boldsymbol{\sigma}\odot\boldsymbol{\epsilon}$, where $\boldsymbol{\epsilon}\sim\mathcal{N}(\boldsymbol{0}, \boldsymbol{1})$).
Each fully-connected hidden layer in the encoder ($f$), decoder ($g$), and auxiliary network ($h$) has a Swish activation function. No activation function is applied to the output layers $\boldsymbol{\mu}$, $\boldsymbol{\sigma}$, $\mathbf{\hat{x}'}$, $\mathbf{\hat{x}''}$, and $\boldsymbol{\hat{\eta}}$. Dashed lines represent a concatenation.}
\label{fig:architecture}
\end{figure}

We use the PyTorch deep learning framework \citep{paszke_pytorch_2019} to implement the CVAE in this study. Figure~\ref{fig:architecture} shows the architecture of the CVAE. The encoder, which comprises three fully-connected hidden layers ($f^{(1)}$ to $f^{(3)}$ in Figure~\ref{fig:architecture}), takes transformed $\sigma_\mathrm{eff}$ spectra as input and yields the mean ($\boldsymbol{\mu}$) and the log-variance ($\log\boldsymbol{\sigma}^2$) parameterizing a normal distribution. Latent representation samples ($\mathbf{z}$) are then drawn from this distribution using the reparameterization trick given by \cite{kingma_auto-encoding_2014}:
\begin{equation}
    \mathbf{z} = \boldsymbol{\mu} + \boldsymbol{\epsilon} \odot \boldsymbol{\sigma},
\end{equation}
where $\boldsymbol{\epsilon}\sim\mathcal{N}(\boldsymbol{0}, \boldsymbol{1})$ is an auxiliary random vector and $\odot$ denotes the Hadamard product. Then, samples of $\mathbf{z}$ are expanded by the decoder ($g^{(1)}$ to $g^{(3)}$ in Figure~\ref{fig:architecture}) back to the original dimensions of the transformed $\sigma_\mathrm{eff}$ spectra. Finally, a parameter decoder ($h^{(1)}$ to $h^{(3)}$ in Figure~\ref{fig:architecture}) simultaneously aims to estimate geomaterial parameters from $\mathbf{z}$ samples. All $f$, $g$ and $h$ functions consist of 32-dimensional fully-connected hidden layers with Swish activation functions \citep{ramachandran_searching_2018}.

In this study, $\mathbf{c}$ represents either none, one, or several of the parameters contained in $\boldsymbol{\eta}$, depending on the available prior information about the mixture. Conditioning is implemented by concatenating $\mathbf{c}$ with $\mathbf{x'}$ and $\mathbf{x''}$ in the input space, and by concatenating $\mathbf{c}$ with $\mathbf{z}$ in the latent space \citep{sohn_learning_2015, molamohammadi_jacobian_2020}.

Training the CVAE requires minimizing the sum of the $D_\textrm{KL}$, a data reconstruction loss ($\mathcal{L}_\mathrm{d}$), and a parameter estimation loss ($\mathcal{L}_\mathrm{p}$). The total training loss $\mathcal{L}$ thus reads
\begin{equation}
\begin{split}
\label{eqn:ppiploss}
    \mathcal{L} = \underbrace{-\frac{1}{2} \sum_{k=1}^K \left(1 + \log{\sigma_k^2} - \mu_k^2 - \sigma_k^2 \right)}_{\textrm{Regularization }(D_\textrm{KL})}
    &+ \underbrace{\frac{1}{J}\sum_{j=1}^J \left[\left(x'_j - \hat{x}'_j \right)^2 + \left(x''_j - \hat{x}''_j \right)^2 \right]}_{\textrm{Data reconstruction }(\mathcal{L}_\mathrm{d})} \\
    &+ \underbrace{\frac{1}{L}\sum_{l=1}^L \left(\eta_l - \hat{\eta}_l \right)^2}_{\textrm{Parameter estimation }(\mathcal{L}_\mathrm{p})},
\end{split}
\end{equation}
where $K$ is the dimensionality of the latent space ($K=8$ unless stated otherwise), $J=32$ is the dimensionality of $\sigma_\mathrm{eff}$, and $L=8$ is the dimensionality of $\boldsymbol{\eta}$. Equation \ref{eqn:ppiploss} is minimized using the Adam stochastic gradient descent algorithm \citep{kingma_adam_2014} using a batch size of 32 and an initial learning rate of $10^{-3}$.

\subsection{Global sensitivity indices}
Sensitivity analysis aims to understand how variations in model inputs affect the model's response and to quantify the relative importance of each input \citep{campolongo_screening_2011}. Here, we define the model inputs as $\boldsymbol{\eta}$, in which each parameter is normalized to the closed unit range $[0, 1]$ to ensure equal relative importance of the model parameters. The model's response is the data reconstruction term in Equation~\ref{eqn:ppiploss}. This choice of model response reflects the inverse modeling objective function when the goal is to estimate geomaterial properties from IP data. \cite{molamohammadi_jacobian_2020} show that global sensitivity indices (SI) can be estimated using the Jacobian of CVAE, which is readily available due to the automatic differentiation feature in modern deep learning frameworks. Moreover, they show that Jacobian-based SI are consistent with those obtained through classic sampling-based methods \citep[e.g.,][]{sobol_global_2001}. The global SI are defined by
\begin{equation}
\label{eqn:sensitivity}
    \mathrm{SI} = \frac{1}{N}\sum_n^N\left(\frac{\partial\mathcal{L}_\mathrm{d}^{(n)}}{\partial\boldsymbol{\eta}^{(n)}}\right)^2,
\end{equation}
where $N$ is the length of $\mathcal{D}_\mathrm{t}$.

\subsection{Standardized accuracy}
We use a standardized accuracy (SA) score \citep{langdon_exact_2016} to evaluate the parameter estimation performance of the CVAE model. The SA is
\begin{equation}
\label{eqn:accuracy}
    \mathrm{SA} = 1 - \frac{\mathrm{MAE}}{\mathrm{MAE_{P_0}}}\times 100,
\end{equation}
where $\mathrm{MAE}$ is the mean absolute error of the parameter predictions and $\mathrm{MAE_{P_0}}$ is the mean absolute error of a random predictor. The SA of a random predictor is thus 0~\%, whereas that of a perfect predictor is 100~\%. We use $\mathrm{MAE_{P_0}}=0.\overline{3}$, which is obtained analytically with the method of \cite{langdon_exact_2016}.

\section{Results}

\subsection{Learning curves and reconstruction quality}
\begin{figure}
\centering
\includegraphics[]{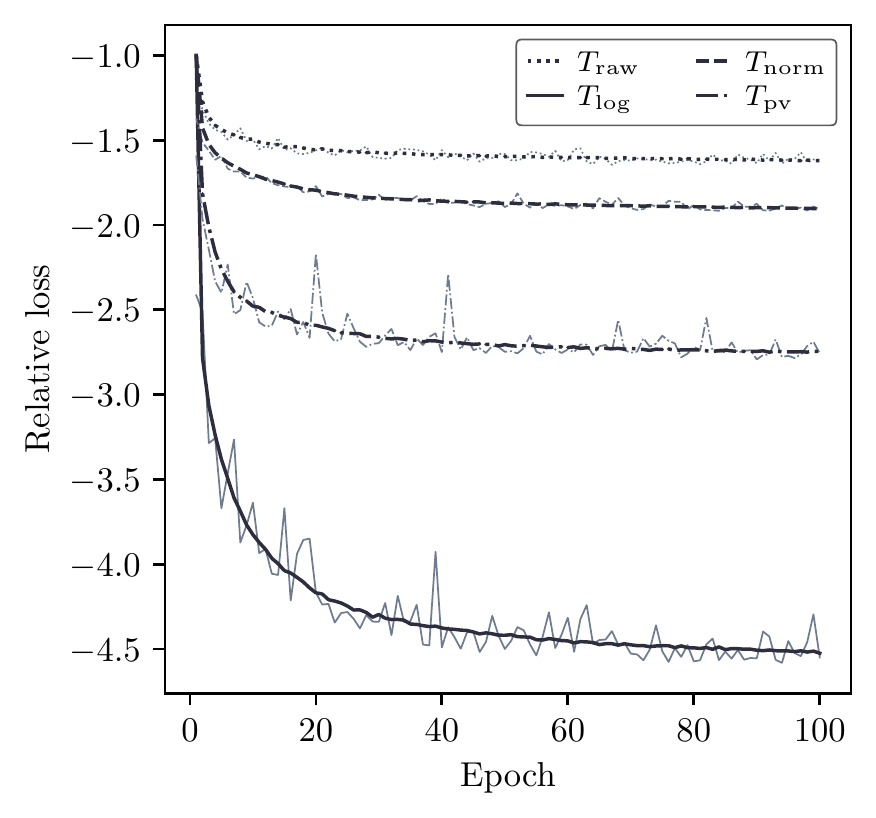}
\caption{CVAE loss over 100 training epochs. For each data transformation $T$, the loss is relative to the loss of the first epoch. Black lines are the training losses, whereas grey lines are the validation losses.}
\label{fig:learning-curves}
\end{figure}

Figure~\ref{fig:learning-curves} shows the evolution of training and validation $\mathcal{L}$ values over 100 epochs. A training epoch consists of optimizing the CVAE weights with forward and backward passes on every $\sigma_\mathrm{eff}$ spectra contained in $\mathcal{D}_\mathrm{t}$. After each training epoch, the CVAE is applied to $\mathcal{D}_\mathrm{v}$ to ensure that the model is not overfitting $\mathcal{D}_\mathrm{t}$.

\begin{figure}
\centering
\includegraphics[]{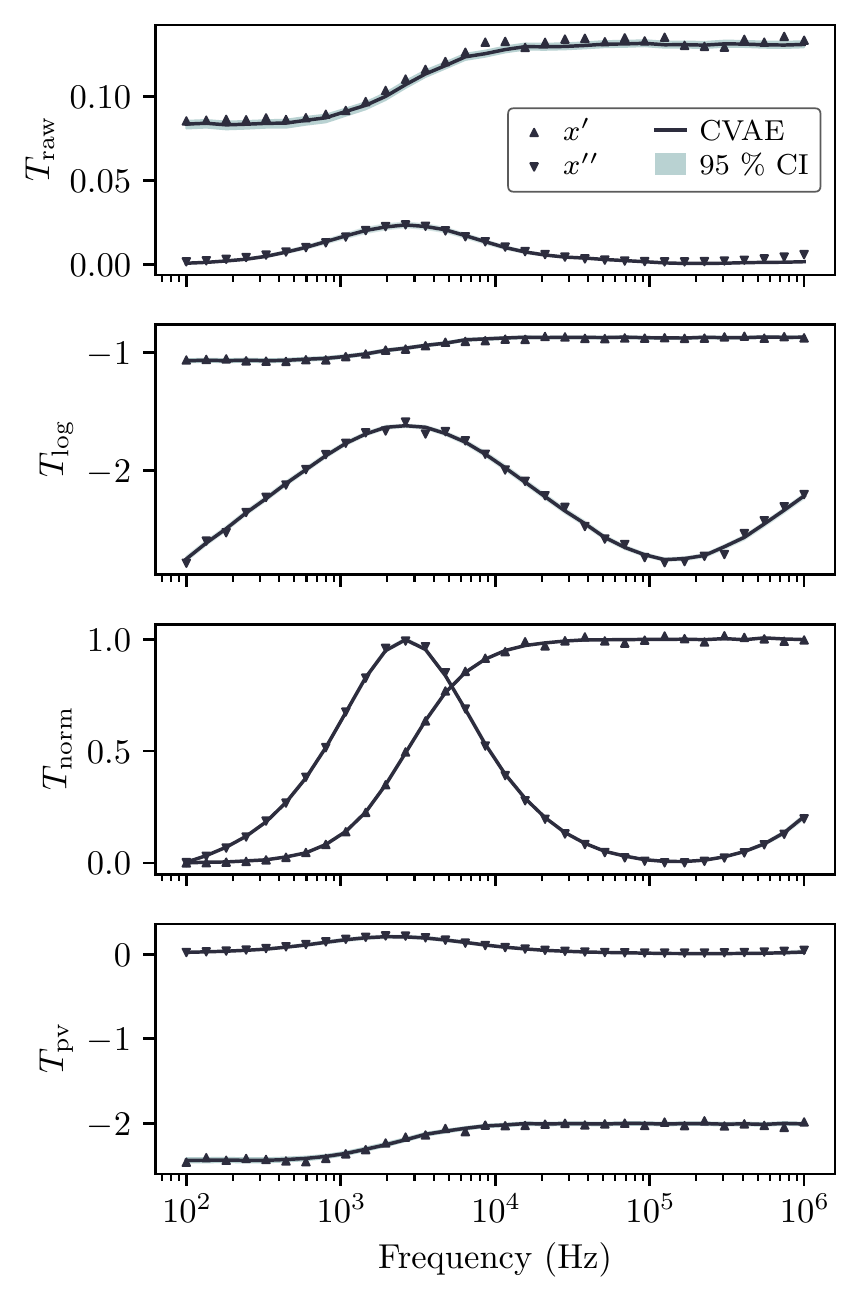}
\caption{CVAE reconstruction results and 95~\% confidence interval (CI) of the test mixture spectra using the four transformations. Table~\ref{tab:parameter-bounds} summarizes the mixture's parameters. One percent additive Gaussian white noise contaminates the data.}
\label{fig:default-mixture-fits}
\end{figure}

In Figure~\ref{fig:learning-curves}, $\mathcal{L}$ values are normalized by the value of the first epoch. This normalization is applied to evaluate the effect of data transformations on the relative change in $\mathcal{L}$ as the CVAE model is optimized. For example, using $T_\mathrm{log}$, the minimum $\mathcal{L}$ values relative to the first epoch are five times smaller. With $T_\mathrm{raw}$, the optimal CVAE $\mathcal{L}$ values are only about 1.5 times smaller than $\mathcal{L}$ of the first epoch. It is evident from Figure~\ref{fig:learning-curves} that optimizing the CVAE requires fewer training epochs with $T_\mathrm{raw}$ than it does with $T_\mathrm{log}$. Qualitatively, the CVAE reconstruction quality for the test mixture is good for all $T$ (Figure~\ref{fig:default-mixture-fits}).

\subsection{Dimension reduction}
\begin{figure}
\centering
\includegraphics[]{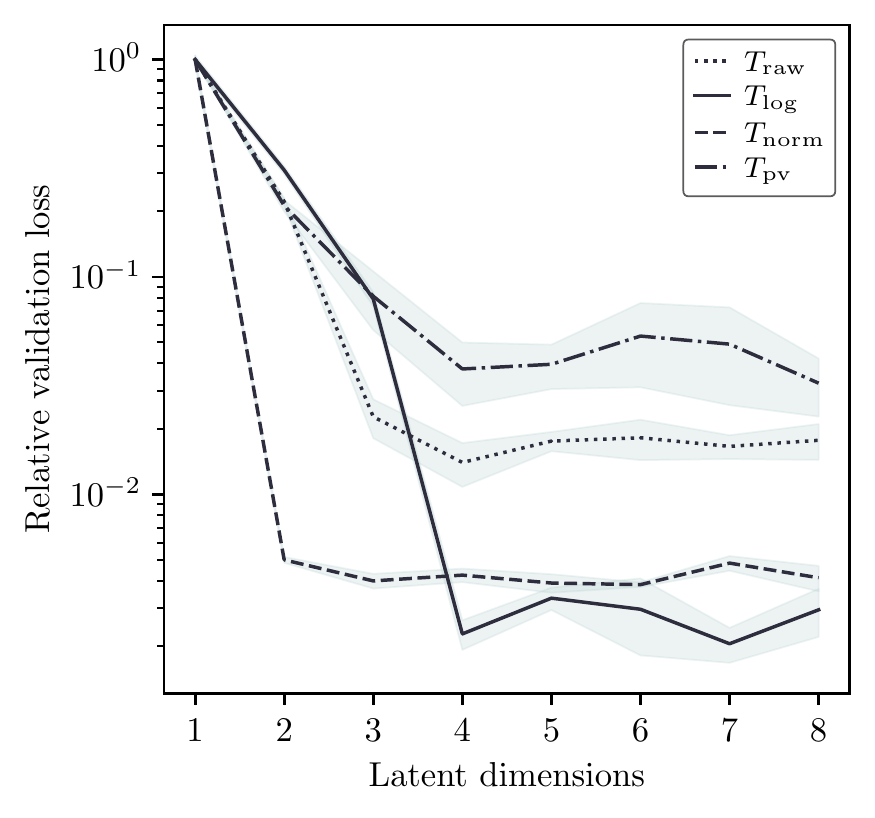}
\caption{Relative data reconstruction loss of the validation data set using CVAE latent distribution dimensions ranging from one to eight. For each $T$, we normalize the loss by its maximum value. The values (lines) and uncertainties (shaded areas) are the mean and standard deviation obtained by repeating the experiment five times.}
\label{fig:latent-dim-MSE}
\end{figure}

Eight petrophysical properties $(a_\mathrm{i}, \phi_\mathrm{i}, D_\mathrm{i}, \sigma_\mathrm{i}, \epsilon_\mathrm{i}, D_\mathrm{h}, \sigma_\mathrm{h}, \epsilon_\mathrm{h})$ govern the PPIP model. However, a dimension reduction experiment can reveal that the PPIP model is compressible to less than eight independent parameters. To conduct this experiment, we train the CVAE on $\mathcal{D}_\mathrm{t}$ using varying latent distribution dimensions ($K\in[1\ldotp \ldotp 8]$), and subsequently apply each model to $\mathcal{D}_\mathrm{v}$. Figure~\ref{fig:latent-dim-MSE} shows the reconstruction error as a function of latent space dimensions for all four data transformations.

With $K=1$, the reconstruction error is maximal, and the CVAE cannot accurately fit the complex conductivity spectra for any $T$. Using $T_\mathrm{norm}$, the total reconstruction error diminishes from $K=1$ to $K=2$, whereafter it stabilizes at a minimum value for $k \in [2\ldotp \ldotp 8]$. Using $T_\mathrm{raw}$, a minimum reconstruction error is obtained with $k \in [3\ldotp \ldotp 8]$. Using $T_\mathrm{log}$ and $T_\mathrm{pv}$, the CVAE model requires at least $K=4$ to fit PPIP spectra with a minimum data reconstruction error.

The relative variations and shape of the error curves shown in Figure~\ref{fig:latent-dim-MSE} indicate that increasing the latent distribution dimensions past $K=4$ does not improve the data reconstruction quality for any $T$. These results suggest that four independent parameters are sufficient to reconstruct $\sigma_\mathrm{eff}$. Thus, only half of the PPIP model parameters may carry meaningful information. Two reasons may explain this discrepancy. First, some of the PPIP model parameters may be strongly correlated, as is the case with Cole-Cole model parameters \citep[e.g.,][]{berube_bayesian_2017}. In that case, two parameters with distinct petrophysical meanings could act as a single parameter through linear combination. Second, variations in some of the PPIP model parameters could have a negligible impact on the overall shape of $\sigma_\mathrm{eff}$ if their sensitivity relative to other model parameters is too low.

For the remainder of the experiments, we fix $K=8$ because the analytical formulation of $\sigma_\mathrm{eff}$ has eight parameters. Setting $K=8$ ensures that the latent space does not act as an information bottleneck in contrast with the analytical formulation.

\subsection{Sensitivity analysis}
\begin{figure}
\centering
\includegraphics[]{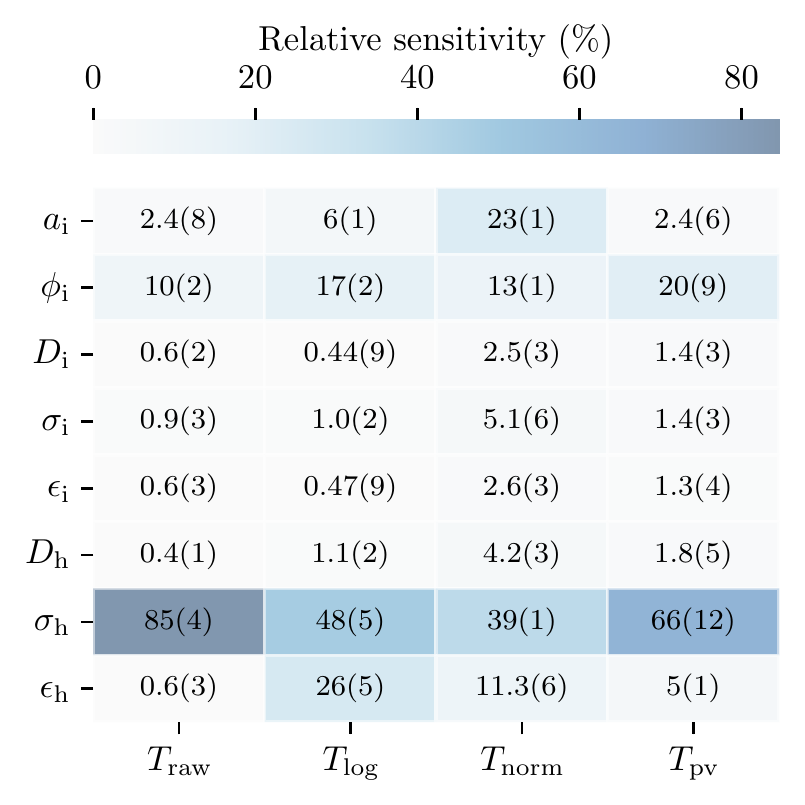}
\caption{Jacobian-based global parameter SI of the PPIP model. For each data transformation $T$, we normalize the rows to sum up to 100~\% to highlight the parameters' relative importances. The values and uncertainties are the mean and standard deviation of five experiment repetitions.}
\label{fig:sensitivity-indices}
\end{figure}

The dimension reduction experiment results indicate that a minimum of four dimensions are required to encode PPIP model data. However, it is unclear which petrophysical parameters could correspond to each dimension and, more specifically, in what order of importance. To answer this question, we conduct a global sensitivity analysis by conditioning the model with $\mathbf{c}=\boldsymbol{\eta}$ and by computing the SI of each petrophysical parameter using the CVAE's Jacobian. Figure~\ref{fig:sensitivity-indices} shows the global SI of the PPIP model for each transformation $T$. The SI are normalized to add up to 100~\% row-wise to highlight each parameter's relative importance.

For any $T$, the most crucial parameter is the host phase's intrinsic conductivity. The diffusion coefficients of both the host and inclusion phases have minor importance compared with other parameters. However, the model's relative sensitivity to the remaining parameters depends on $T$.

Using $T_\mathrm{raw}$, the PPIP model's sensitivity is dominated by $\sigma_\mathrm{h}$ ($\mathrm{SI} = 85\pm 4~\%$), followed to a lesser extent by $\phi_\mathrm{i}$ ($\mathrm{SI} = 10\pm 2~\%$), and $a_\mathrm{i}$ ($\mathrm{SI} = 2.4\pm 8~\%$). Using $T_\mathrm{log}$ significantly alters the relative SI of the PPIP model parameters.
First, it reduces the model's relative sensitivity to $\sigma_\mathrm{h}$ ($\mathrm{SI} = 48\pm 5~\%$) while increasing its sensitivity to $\epsilon_\mathrm{h}$ ($\mathrm{SI} = 26\pm 5~\%$), $\phi_\mathrm{i}$ ($\mathrm{SI} = 17\pm 2~\%$), and $a_\mathrm{i}$ ($\mathrm{SI} = 6\pm 1~\%$). Using $T_\mathrm{pv}$ yields similar results to using $T_\mathrm{log}$, with the exception of a lower relative SI for $\epsilon_\mathrm{h}$ ($\mathrm{SI} = 5\pm 1~\%$). Using $T_\mathrm{norm}$ yields the most evenly distributed SI across the PPIP model parameters when compared to other $T$.

The host phase's intrinsic conductivity consistently has the highest importance because this parameter controls the amplitude of the complex conductivity's in-phase component. It is interesting to note that $\sigma_\mathrm{eff}$ spectra transformed by $T_\mathrm{log}$ are more sensitive to the $\phi_\mathrm{i}$ than when using $T_\mathrm{raw}$. Moreover, the SI values imply that using $T_\mathrm{log}$ augments the relative sensitivity of the PPIP model to $a_\mathrm{i}$. From a parameter estimation perspective, our results suggest that $\phi_\mathrm{i}$ and $\sigma_\mathrm{h}$ should be recoverable from $\sigma_\mathrm{eff}$ for any $T$. However, other parameters such as $D_\mathrm{i}$, $\sigma_\mathrm{i}$, $\epsilon_\mathrm{i}$, and $D_\mathrm{h}$ may have ill-defined solutions due to their negligible relative importance for fitting the PPIP model to $\sigma_\mathrm{eff}$. These results are consistent with evidence that a minimum of four latent dimensions are needed to minimize the PPIP model data reconstruction error.

\subsection{Parameter estimation}
Estimating $\boldsymbol{\eta}$ using $\sigma_\mathrm{eff}$ is desirable for non-invasive characterization of the subsurface with the IP method. In this section, we analyze the PPIP model's parameter estimation SA in the case where the model is not conditioned (unconstrained inversion) and in the case where the model is conditioned by specific parameters (constrained inversion). For each experiment, the CVAE is trained on $\mathcal{D}_\mathrm{t}$ and tested on $\mathcal{D}_\mathrm{v}$.

\subsubsection{No conditioning}
\begin{figure}
\centering
\includegraphics[]{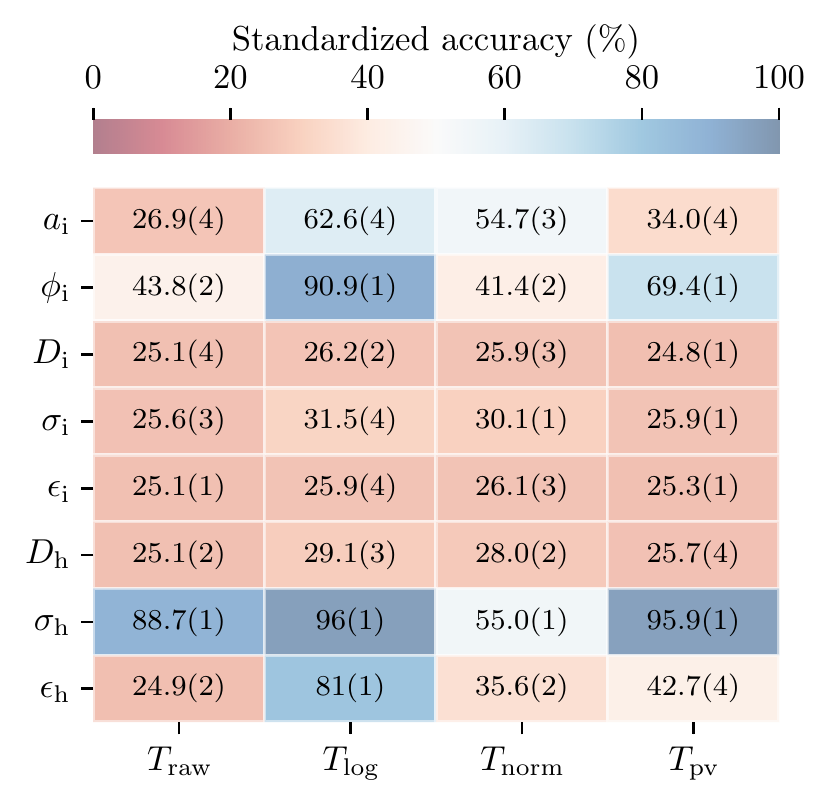}
\caption{Standardized parameter estimation accuracy for each applied data transformation $T$ and without conditioning. The values and uncertainties are the mean and standard deviation obtained by repeating the experiment five times.}
\label{fig:parameter-estimation}
\end{figure}

Figure~\ref{fig:parameter-estimation} summarizes the SA scores averaged over $\mathcal{D}_\mathrm{v}$ with respect to the four data transformations used in this study and when none of the parameters are constrained. Using $T_\mathrm{raw}$, only the most sensitive PPIP model parameter ($\sigma_\mathrm{h}$) can be estimated with a relatively high SA of $88.7\pm 0.1~\%$.
We observe that $D_\mathrm{i}$, $\sigma_\mathrm{i}$, $\epsilon_\mathrm{i}$, $D_\mathrm{h}$, and $\epsilon_\mathrm{h}$ have the lowest SA measures with approximately 25~\%. Furthermore, it is not possible to estimate $\phi_\mathrm{i}$ ($\mathrm{SA} = 43.8\pm 0.2~\%$) and $a_\mathrm{i}$ ($\mathrm{SA} = 26.9\pm 0.4~\%$) with SA values above 50~\%.

As evidenced in Figure~\ref{fig:parameter-estimation}, SA measures increase by approximately 35~\% for $a_\mathrm{i}$ ($\mathrm{SA} = 62.6\pm 0.4~\%$), by approximately 45~\% for $\phi_\mathrm{i}$ ($\mathrm{SA} = 90.9\pm 0.1~\%$), and by approximately 55~\% for $\epsilon_\mathrm{h}$ ($\mathrm{SA} = 81\pm 1~\%$) when applying $T_\mathrm{log}$ to $\sigma_\mathrm{eff}$. This observation concords with the change in SI of these parameters when using $T_\mathrm{log}$. Still using $T_\mathrm{log}$, the SA measures of all other parameters remain fairly unchanged in comparison with those obtained with $T_\mathrm{raw}$.

Applying $T_\mathrm{norm}$ to $\sigma_\mathrm{eff}$ yields the lowest SA measures out of all $T$ (Figure~\ref{fig:parameter-estimation}). With normalization, the SA measure of $\sigma_\mathrm{h}$ is only $55.0\pm 0.1~\%$, despite this parameter being the most important one according to sensitivity analysis. Moreover, in comparison with $T_\mathrm{raw}$, the SA measures for $\phi_\mathrm{i}$ and $\epsilon_\mathrm{h}$ are considerably lower when using $T_\mathrm{norm}$. The only benefit to using $T_\mathrm{norm}$ in contrast with $T_\mathrm{raw}$ is a slight increase in the SA measure of $a_\mathrm{i}$, which is also concordant with the sensitivity analysis results.

Applying $T_\mathrm{pv}$ to $\sigma_\mathrm{eff}$ yields SA measures that are similar but slightly inferior to those obtained with $T_\mathrm{log}$.

To summarize, $\phi_\mathrm{i}$, the $\sigma_\mathrm{h}$, and $\epsilon_\mathrm{h}$ can be estimated from $\sigma_\mathrm{eff}$ with close to 100~\% SA when using either $T_\mathrm{log}$ or $T_\mathrm{pv}$. Estimating $a_\mathrm{i}$ is possible when $\sigma_\mathrm{eff}$ has been subjected to $T_\mathrm{log}$, $T_\mathrm{norm}$, or $T_\mathrm{pv}$. However, the SA measure for this parameter is only approximately 60~\%. We recommend avoiding $T_\mathrm{raw}$ or $T_\mathrm{norm}$ because these transformations yield significantly lower SA scores across the board. None of the applied $T$ provide acceptable SA scores for $D_\mathrm{i}$, $\sigma_\mathrm{i}$, $\epsilon_\mathrm{i}$, and $D_\mathrm{h}$ in the unconstrained case.

\subsubsection{With conditioning}
When one or several parameters from $\boldsymbol{\eta}$ condition the CVAE during training, the parameter decoder acts as a constrained parameter estimator. We now compare the parameter estimation SA measures obtained with various conditioning scenarios that may arise in practical applications of the IP method. The experiment is performed only on $T_\mathrm{log}\colon \sigma_\mathrm{eff}$ because it consistently yields the best average SA measures (Figure~\ref{fig:parameter-estimation}). The parameter estimation scenarios (A--G) are:

\begin{enumerate}[label=Scenario \Alph*., leftmargin=*]
    \item This scenario aims at estimating all petrophysical parameters without prior information. This scenario is the general unconstrained inversion case for which no information is available about the subsurface.
    \item This scenario aims at estimating the electrochemical and geometric parameters of the inclusions, given the electrochemical properties of the host. This scenario reflects applications of the IP method to detect conductive minerals in a host rock or soil with well-known properties.
    \item This scenario aims at estimating the geometry of the inclusions and electrochemical properties of the host, given the electrochemical properties of the inclusions. This scenario arises in mineral exploration, where the nature of the target metallic mineral is known, but its geometrical properties and the host's nature may vary.
    \item This scenario aims at estimating  $a_\mathrm{i}$ and $\phi_\mathrm{i}$, given all electrochemical properties of the subsurface. This scenario arises in mineral exploration, where the natures of both inclusion and host phases are known, and the task is to estimate the target mineral resources.
    \item This scenario aims at estimating the electrochemical parameters of both host and inclusion phases, given $a_\mathrm{i}$ and $\phi_\mathrm{i}$. This scenario may arise in laboratory experiments using synthetic mixtures.
    \item This scenario aims at estimating the electrochemical parameters of the host material, given the geometrical and electrochemical properties of the inclusion phase. This scenario may arise in laboratory experiments that monitor host phase properties exhibiting a time dependence.
    \item This scenario aims at estimating the electrochemical properties of the inclusion phase, given the inclusions' geometrical parameters and the host material's electrochemical properties. This scenario may arise in applications aiming to identify the nature of the conductive inclusions in well-known host materials.
\end{enumerate}

\begin{figure}
\centering
\includegraphics[]{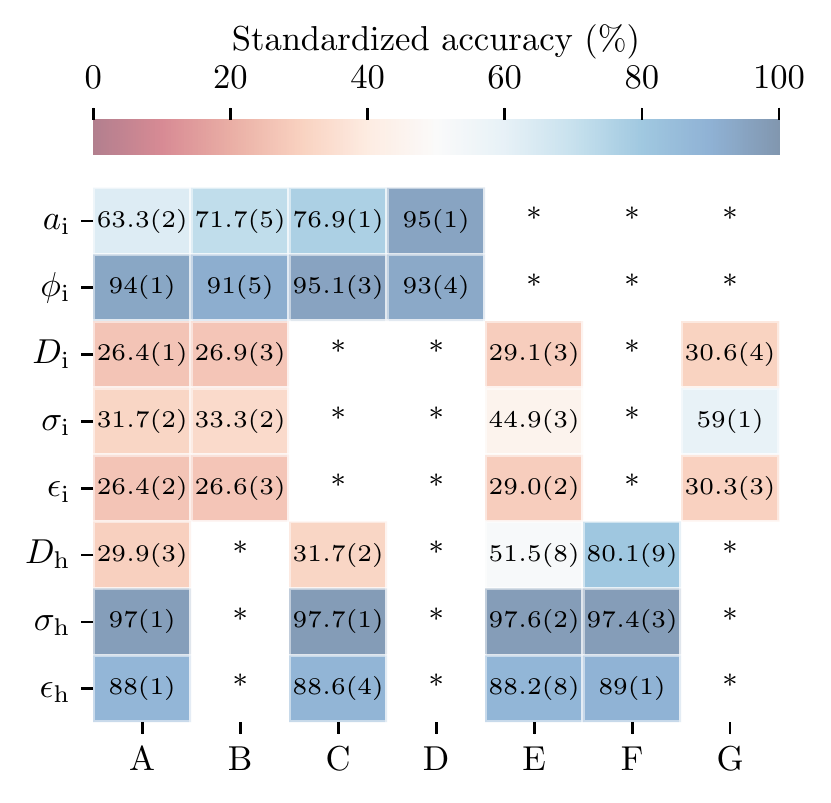}
\caption{Standardized parameter estimation accuracy using $T_\mathrm{log}$ and various conditioning scenarios (A--G). Asterisks (*) mark the constrained parameters. A: No conditions. B: Host electrochemical properties are known. C: Inclusion electrochemical properties are known. D: Electrochemical properties of inclusions and host are known. E: Geometrical properties of the inclusions are known. F: Electrochemical and geometrical properties of the inclusions are known. G: Inclusion phase geometry and host electrochemical properties are known.}
\label{fig:cond-parameter-estimation-log}
\end{figure}

Figure~\ref{fig:cond-parameter-estimation-log} provides the SA scores of each PPIP parameter estimated from $\sigma_\mathrm{eff}$ using conditioning scenarios A--G. Small disparities between the SA scores provided in Figure~\ref{fig:parameter-estimation} and Figure~\ref{fig:cond-parameter-estimation-log}A are caused by the stochastic nature of the CVAE model when repeating the experiments.

Conditioning the CVAE with the host material's electrochemical properties improves the parameter estimation SA scores of unknown parameters (Figure~\ref{fig:cond-parameter-estimation-log}B). In addition, knowing the electrochemical properties of the inclusion phase increases the SA score of $a_\mathrm{i}$ by approximately 14~\% ($76.9\pm 0.1$ \%) in comparison with the unconstrained scenario (Figure~\ref{fig:cond-parameter-estimation-log}C).

Prior knowledge of both inclusion and host phases' electrochemical properties improves the SA measures of the inclusions' geometrical properties. If all geomaterial electrochemical properties are constrained, the SA of $a_\mathrm{i}$ is $95\pm 1$ \%, which marks an improvement of more than 30~\% over the unconstrained inversion case (Figure~\ref{fig:cond-parameter-estimation-log}D).

$\sigma_\mathrm{i}$, which is not recoverable in the unconstrained case, can be estimated with a SA score of $44.9\pm0.3$ \% when $a_\mathrm{i}$ and $\phi_\mathrm{i}$ are known (Figure~\ref{fig:cond-parameter-estimation-log}E), marking an improvement of approximately 13\% in comparison with the unconstrained inversion. The SA score for $\sigma_\mathrm{i}$ is further increased to $59\pm1$ \% if the electrochemical properties of the inclusions phase are the only unknown parameters (Figure~\ref{fig:cond-parameter-estimation-log}G).

As a general rule, it is impossible to estimate $D_\mathrm{i}$ and $\epsilon_\mathrm{i}$ with SA measures above 50~\%, regardless of the available prior information about the geomaterial. This observation is consistent with the PPIP model's lack of sensitivity to these parameters. However, we note that the SA score of $D_\mathrm{h}$ reaches $80.1\pm0.9$ \% when all geometrical and electrochemical parameters of the inclusion phase are constrained (Figure~\ref{fig:cond-parameter-estimation-log}F).

\subsection{Effective parameter space}
This section inspects the latent distribution learned by training the CVAE on PPIP model data. The data decoder can sample this distribution to generate synthetic $\sigma_\mathrm{eff}$ spectra, whereas sampling the parameter decoder yields the corresponding $\boldsymbol{\eta}$. We first compare the parameter space learned by the CVAE when $T_\mathrm{raw}$ and $T_\mathrm{log}$ are applied to $\sigma_\mathrm{eff}$. Then, we demonstrate that conditioning the model with $a_\mathrm{i}$ has a smoothing effect on the effective parameter space and that this constraint mitigates the inter-parameter correlations.

\subsubsection{No conditioning}
\begin{figure}
\centering
\includegraphics[]{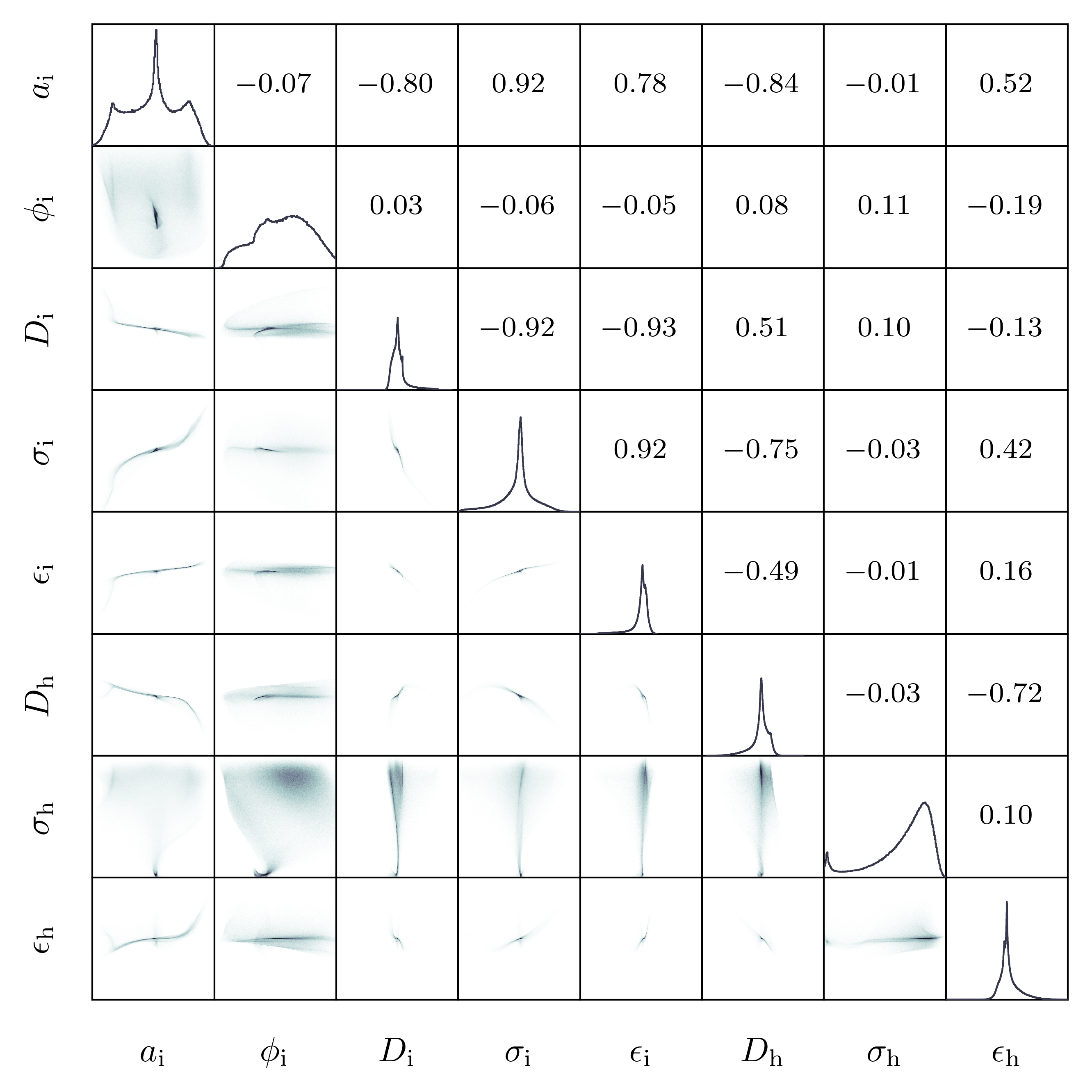}
\caption{Pairwise distribution plots and Pearson correlation coefficients of 1~000~000 PPIP model parameter sets sampled from the CVAE parameter decoder when $T_\mathrm{raw}$ is applied. Darker shades correspond to denser regions of the parameter space. The edges of each distribution plot correspond to the parameter bounds (Table~\ref{tab:parameter-bounds}).}
\label{fig:parameter-posterior-raw}
\end{figure}

Figure~\ref{fig:parameter-posterior-raw} shows the distribution plots of the PPIP parameter values learned by the CVAE when no transformation is applied. Dark areas in the distribution plots correspond to PPIP model parameter space regions that are most useful to fully reconstruct $\mathcal{D}_\mathrm{t}$. Figure~\ref{fig:parameter-posterior-raw} is thus a representation of the model's sensitivity distribution throughout the parameter space.

The results in Figure~\ref{fig:parameter-posterior-raw} are concordant with the global sensitivity analysis and parameter estimation limitations of the PPIP model. The most sensitive parameters, which are recoverable with high accuracy, have distributions that cover most of the area of their prior distributions (Table~\ref{tab:parameter-bounds}). Therefore, to fit all $\sigma_\mathrm{eff}$ contained in $\mathcal{D}_\mathrm{t}$, the CVAE has learned that $\sigma_\mathrm{h}$, $\phi_\mathrm{i}$, $a_\mathrm{i}$, and $\sigma_\mathrm{i}$ must vary across most of the available prior parameter space.
Contrastingly, the distributions of insensitive parameters $D_\mathrm{i}$, $\sigma_\mathrm{i}$, $\epsilon_\mathrm{i}$, and $D_\mathrm{h}$ have small standard deviations and are centered on their priors' mean values. Therefore, the CVAE has learned that using the mean value of these parameters is sufficient to reconstruct all data in $\mathcal{D}_\mathrm{t}$. This result implies that a PPIP model with fixed mean values of $D_\mathrm{i}$, $\sigma_\mathrm{i}$, $\epsilon_\mathrm{i}$, and $D_\mathrm{h}$ will fit any $\sigma_\mathrm{eff}$ spectra generated within the boundaries given in Table~\ref{tab:parameter-bounds} with no loss of reconstruction quality.

Overall, we note that the $T_\mathrm{raw}$ PPIP model parameter space features complex patterns, a multitude of local high-density areas, strong inter-parameter correlation (e.g., Pearson $r=0.84$ for $a_\mathrm{i}$ and $\sigma_\mathrm{i}$), and accumulation near the upper boundary of $\sigma_\mathrm{h}$. These features could explain why the parameter estimation accuracy using $T_\mathrm{raw}$ is relatively low as gradient-based optimization algorithms may get stuck in local minima or yield ill-defined results due to inter-parameter correlation.

\begin{figure}
\centering
\includegraphics[]{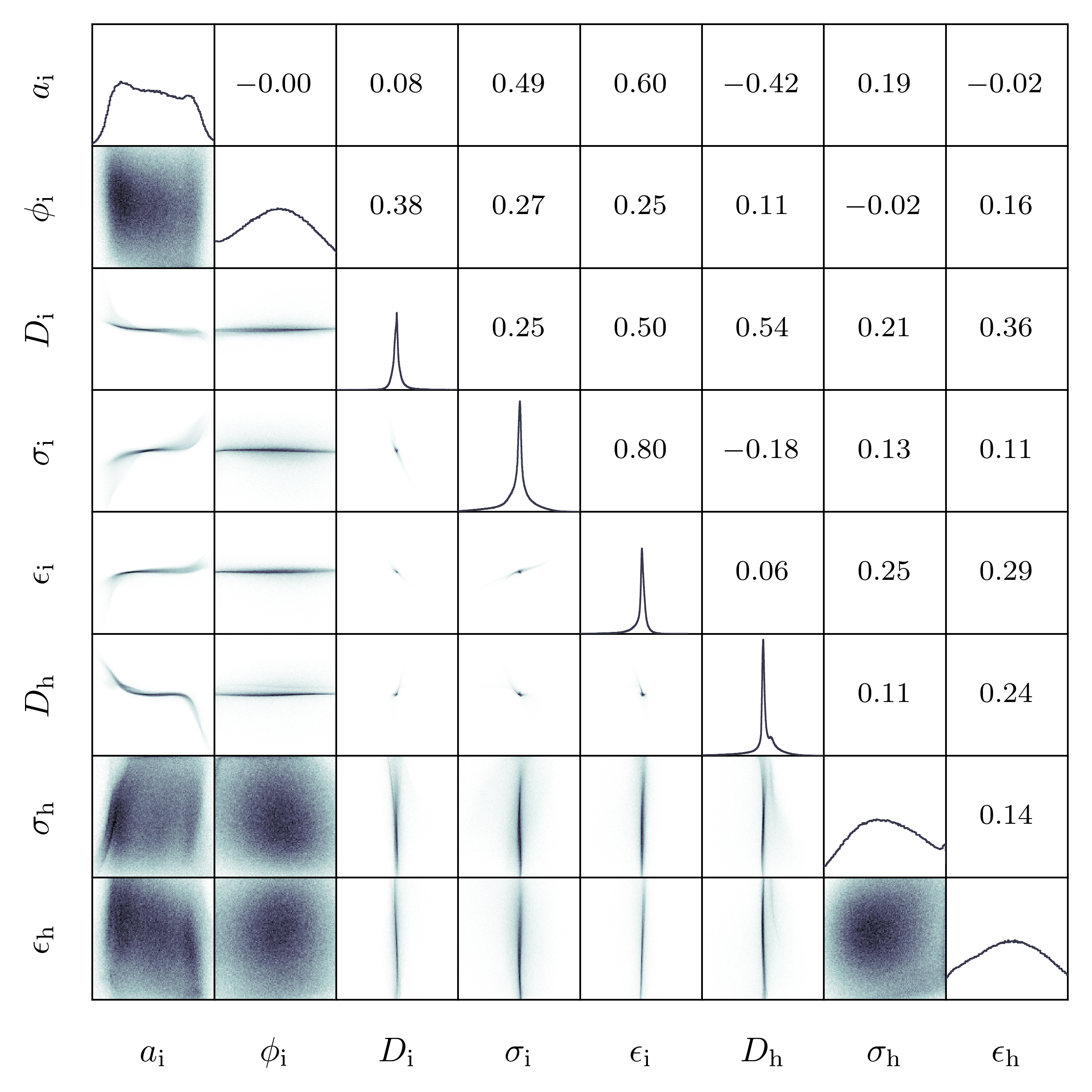}
\caption{Pairwise distribution plots and Pearson correlation coefficients of 1~000~000 PPIP model parameter sets sampled from the CVAE parameter decoder when $T_\mathrm{log}$ is applied. Darker shades correspond to denser regions of the parameter space. The edges of each distribution plot correspond to the parameter bounds (Table~\ref{tab:parameter-bounds}).}
\label{fig:parameter-posterior-log}
\end{figure}

Figure~\ref{fig:parameter-posterior-log} shows the effective parameter space learned by the CVAE when $T_\mathrm{log}$ is applied to $\sigma_\mathrm{eff}$. Under $T_\mathrm{log}$, the parameter space appears smoother than it does with $T_\mathrm{raw}$. This is most evident when inspecting the bivariate distribution plots of $a_\mathrm{i}$, $\phi_\mathrm{i}$, $\sigma_\mathrm{h}$, and $\epsilon_\mathrm{h}$, for which the learned effective parameter space covers most of the prior space (Table~\ref{tab:parameter-bounds}). These four parameters also have the highest sensitivity indices (Figure~\ref{fig:sensitivity-indices}) and can be estimated accurately from $T_\mathrm{log}\colon\sigma_\mathrm{eff}$ (Figure~\ref{fig:parameter-estimation}). Overall, in comparison with $T_\mathrm{raw}\colon\sigma_\mathrm{eff}$, the effective parameter space of $T_\mathrm{log}\colon\sigma_\mathrm{eff}$ has smaller inter-parameter correlations (e.g., Pearson $r=0.61$ for $a_\mathrm{i}$ and $\sigma_\mathrm{i}$), no accumulation near the boundaries, and less local high-density areas. These characteristics may provide an easier optimization landscape for parameter estimation tasks.

\subsubsection{With conditioning}

\begin{figure}
\centering
\includegraphics[]{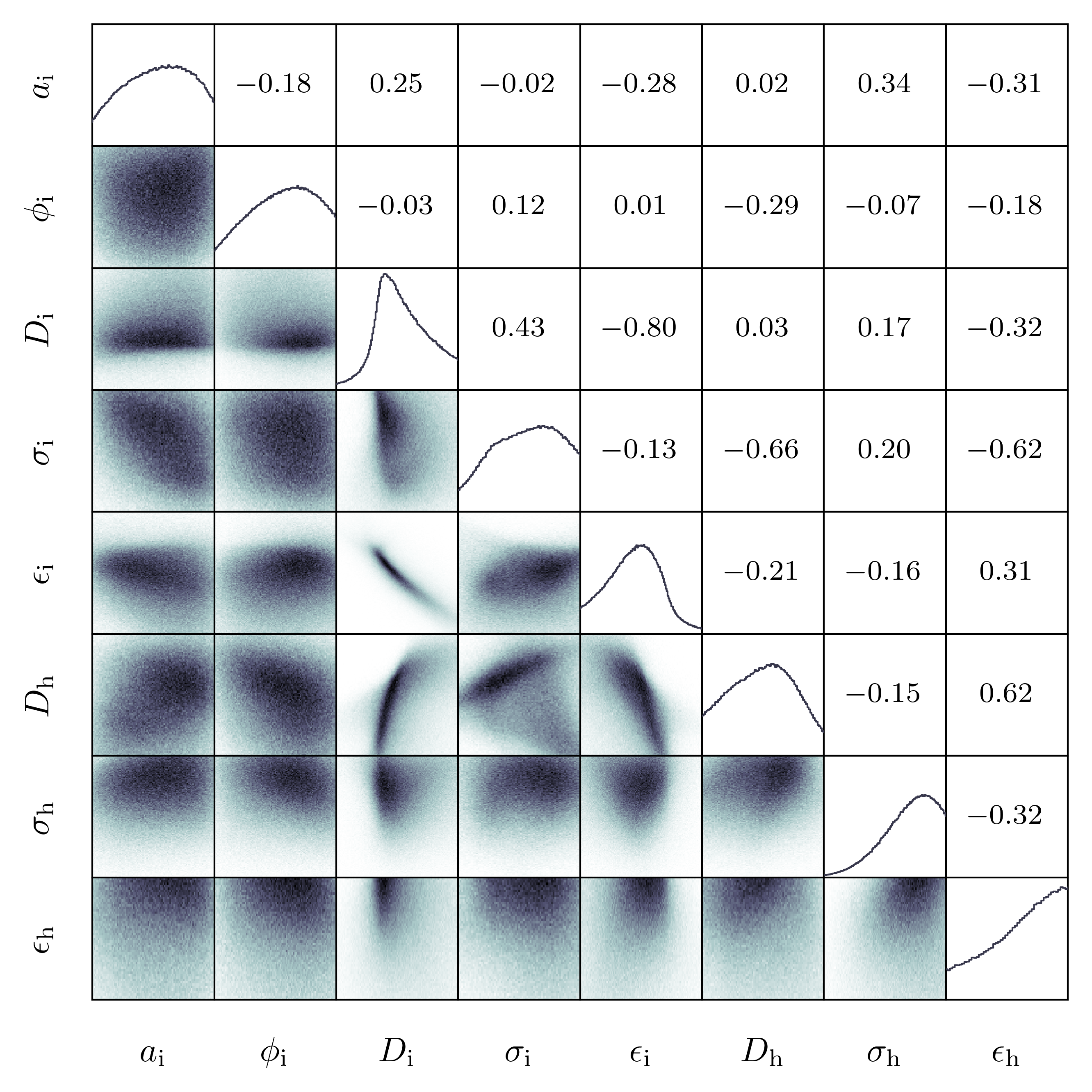}
\caption{Pairwise distribution plots and Pearson correlation coefficients of 1~000~000 PPIP model parameter sets sampled from the CVAE parameter decoder when $T_\mathrm{log}$ is applied and when $a_\mathrm{i}$ is known. Darker shades correspond to denser regions of the parameter space. The edges of each distribution plot correspond to the parameter bounds (Table~\ref{tab:parameter-bounds}).}
\label{fig:parameter-posterior-log-cond0}
\end{figure}

It is evident from Figures~\ref{fig:parameter-posterior-raw}~and~\ref{fig:parameter-posterior-log} that $a_\mathrm{i}$ and $\sigma_\mathrm{i}$ are correlated. Moreover, prior knowledge of $a_\mathrm{i}$ significantly improves the parameter estimation scores for $\sigma_\mathrm{i}$ (Figure~\ref{fig:cond-parameter-estimation-log}). Figure~\ref{fig:parameter-posterior-log-cond0} shows how conditioning the CVAE model with $a_\mathrm{i}$ alters the effective parameter space of $T_\mathrm{log}\colon\sigma_\mathrm{eff}$.

In contrast with the unconditioned parameter space, the $T_\mathrm{log}\colon\sigma_\mathrm{eff}$ parameter space conditioned by $a_\mathrm{i}$ is smoother and more representative of the prior parameter space. Conditioning the model with $a_\mathrm{i}$ also further reduces the overall inter-parameter correlation. In addition, any local high-density regions are more diffuse when $a_i$ is known, which may further simplify the optimization landscape for parameter estimation tasks. Overall, the conditional model fills regions of the parameter space that the unconditional model cannot cover (Figure~\ref{fig:parameter-posterior-log}). This is most evident when comparing the electrochemical parameters of the inclusion phase in Figures~\ref{fig:parameter-posterior-log}~and~\ref{fig:parameter-posterior-log-cond0}.

\section{Conclusions}
IP researchers and practitioners should not use mechanistic IP models to predict the intrinsic petrophysical properties of the subsurface naively because, as with their empirical counterparts, mechanistic models suffer from over-parameterization. This study shows that a generative model of $\sigma_\mathrm{eff}$ is only sensitive to four petrophysical properties. Our data-driven findings contrast with the fact that eight petrophysical properties govern the polarization mechanism, according to the analytical PPIP model formulation. From most to least important, the four sensitive parameters are $\sigma_\mathrm{h}$, $\epsilon_\mathrm{h}$, $\phi_\mathrm{i}$, and $a_\mathrm{i}$. The relatively insensitive parameters are $D_\mathrm{i}$, $\sigma_\mathrm{i}$, $\epsilon_\mathrm{i}$, and $D_\mathrm{h}$.

Over-parameterization has unfortunate implications for interpreting IP data with the PPIP model. Unconstrained inverse modeling of $\sigma_\mathrm{eff}$ can only yield accurate estimations of $\sigma_\mathrm{h}$, $\phi_\mathrm{i}$, $\epsilon_\mathrm{h}$, and to a lesser extent $a_\mathrm{i}$. However, our constrained inverse modeling experiment provides new theoretical implications for metallic mineral discrimination using the IP method (e.g., delineating pyrite zones from pyrrhotite zones). If the precise $\phi_\mathrm{i}$ and $a_\mathrm{i}$ values of a geomaterial are known, as well as the intrinsic electrochemical properties of the host, identification of the inclusions' nature by estimating $\sigma_\mathrm{i}$ from $\sigma_\mathrm{eff}$ is conceivable. Without meeting these conditions, however, identifying the nature of metallic inclusions by fitting the PPIP model to $\sigma_\mathrm{eff}$ data appears impossible.

Applying either $T_\mathrm{log}$ or $T_\mathrm{pv}$ to $\sigma_\mathrm{eff}$ is beneficial to the PPIP model's petrophysical parameter estimation accuracy. In comparison with using $T_\mathrm{raw}$, $T_\mathrm{log}$ alters the relative importance of the PPIP model parameters and provides the best overall parameter estimation accuracy. The untransformed, unconstrained PPIP model parameter space is complex, features many local high-density regions, and suffers from strong inter-parameter correlations (namely between $a_\mathrm{i}$ and $\sigma_\mathrm{i}$). These issues can be mitigated by using $T_\mathrm{log}$ instead of $T_\mathrm{raw}$ when interpreting $\sigma_\mathrm{eff}$ data with the PPIP model.

Diffusion of the charge carriers during relaxation time is an essential component of the metallic minerals polarization mechanism as we currently understand it. Then, how can we explain the generative modeling insensitivity of $\sigma_\mathrm{eff}$ to $D_\mathrm{i}$ and $D_\mathrm{h}$? A lack of generative modeling sensitivity for a certain parameter does not mean that it is useless for the model as a whole. It does mean, however, that this parameter could take on any value, and careful adjustment of the other parameters would still allow a perfect fit between the PPIP model and data. Consequently, insensitive parameters should be systematically constrained using the best available information when interpreting $\sigma_\mathrm{eff}$ data with the PPIP model.

The proposed CVAE framework is applicable to more complex geomaterial mixtures modeled with the PPIP equations or any other mechanistic, empirical or data-driven IP model. Users can even train the CVAE on collections of $\sigma_\mathrm{eff}$ data generated by multiple models, as long as the frequency range is the same. Finally, the CVAE framework is applicable to model the $\sigma_\mathrm{eff}$ of real geomaterials samples. The actual petrophysical parameters are likely unknown for real geomaterials. However, other observations such as modal mineralogy, whole-rock geochemistry, or rock type categories may be available. The CVAE can then be conditioned and applied for sensitivity and parameter estimation analyses concerning those variables.

\begin{acknowledgments}
  The first author acknowledges support from Polytechnique Montréal's new faculty start-up grant program. 
\end{acknowledgments}

\bibliographystyle{seg}  
\bibliography{references}

\end{document}